\documentclass{iau}
\usepackage{graphicx}
\usepackage{natbib}

\title[The central pc-scale region in blazars] 
{The central pc-scale region in blazars: insights from multi-band observations}

\author[Tigran G.\ Arshakian and Vahram Chavushyan]   
{Tigran G.\ Arshakian$^{1,2}$
and Vahram Chavushyan$^{3}$}

\affiliation{$^1$I. Physikalisches Institut, Universit\"at zu K\"oln,  Z\"ulpicher\\ Strasse 77, D-50937 K\"oln, Germany \\ email: {\tt arshakian@ph1.uni-koeln.de} \\[\affilskip]
$^2$Byurakan Astrophysical Obsevatory, Aragatsotn prov.\\ 378433, Armenia \\
$^3$Instituto Nacional de Astrof\'isica, \'Optica y Electr\'onica, \\ Apartado Postal 51 y 216, 72000 Puebla, Pue, M\'exico }

\pubyear{2008}
\volume{xxx}  
\pagerange{119--126}
\jname{Title of your IAU Symposium}
\editors{A.C. Editor, B.D. Editor \& C.E. Editor, eds.}
\begin{document}

\maketitle

\begin{abstract}
The empirical relations in the black hole-accretion disk-relativistic jet system and physical 
processes behind these relations are still poorly understood, partly because they operate close to the black hole within the central light year. Very long baseline array (VLBA) provides
unparalleled resolution at 15 GHz with which to observe the jet components at
sub-milliarcsecond scales, corresponding to sub-pc-scales for local blazars. 
We discuss the jet inner structure of blazars, location and radiation mechanisms operating in the innermost parsec-scale region of blazars, and evidence for jet-excited broad-line region (BLR) ouflowing downstream the jet. Outflowing BLR can provide necessary conditions for production of high energy emission along the jet between the base of the jet and the BLR and far beyond the BLR as evidenced by recent observations. Flat spectrum quasars and low synchrotron peaked sources are the most likely objects to host the outfllowing BLR.
From the $\gamma$-ray absorption arguments, we propose that the jet-excited region of the outflowing BLR in quasars is small and/or gas filling factor is low, and that the orientation and opening angle of the outflowing BLR can lead to relevant $\gamma$-ray absorption features observed in quasars.
%

\end{abstract}

\firstsection 
\section{Inner jet structure: radio core and recolimation shock}
Relativistic collimated outflows of plasma material (jet) are observed in radio domain with VLBA in nearby radio galaxies (RG), BL Lacertae objects (BL Lac) and flat spectrum radio quasars (FSRQ). High resolution radio observations at 15 GHz and 43 GHz show inhomogeneities in the jet plasma, which appear as compact bright knots ranging from subparsec- to pc-scales. The brightest knot is identified as the radio core of the jet which is likely to be the region where the optical depth becomes $\approx 1$. Series of quasi-stationary knots and knots moving downstream the jet are often identified from the VLBA monitoring of blazars. The first stationary component from the radio core is located at submilliarcsecond scales which translates to a sub-parsec distance for nearby RGs 3C 390.3, 3C 120, BL Lacertae and M 87, and FSRQs 3C 273 and 3C 345.3 \citep[e.g.,][]{arshakian10,tavares10,cohen14,cheung07,jorstad05,schinzel12}. This stationary component appears to be a common feature in blazars and it is likely to be a standing recollimation shock (RCS), which is evident from HD/MHD simulations of jets \citep[e.g.,][]{gomez95, krause01}. The RCS indicates the end of the magnetically dominated zone where the continuous jet flow is accelerated and the location where a strong recollimation shock forms and moves downstream the jet with superluminal speed. 

It is generally excepted that the bulk of total synchrotron and polarized emission of the pc-scale jet from radio to X-ray frequencies are generated in the innermost part of the jet. The synchrotron radio and polarized emission is emitted in the radio core from relativistic electrons in the jet. The ejection of superluminal components from the RCS in RGs (3C 390.3 and 3C 120) are correlated with optical continuum flares \citep{arshakian10,tavares10}, suggesting that the source of variable optical emission is non-thermal and located between the radio core and the RCS. 
Variable synchrotron optical flare and high levels of polarization are 
generated around the radio core in quasars \citep{darcangelo07} and in the innermost $\approx0.2$ pc region of the jet in BL Lacs \citep{marscher08}. Evidence for $\gamma$-ray emission generated in the innermost part of the jet is discussed in the next section.

\section{Location of $\gamma$-ray emission in the jet}
About 80\,\% of blazars from the MOJAVE (Monitoring of Jets in AGN
with VLBA Experiments) radio selected sample are $\gamma$-ray sources detected with \emph{Fermi} Large Area Telescope \citep{lister09}. SED of blazars from radio band to high energies are generally well understood. The jet model with only synchrotron self-Compton (SSC) component is able to fit the SED of high synchrotron peaked (HSP) sources, while the SSC and external inverse-Compton (EC) are required to fit the low and high energy bumps of FSRQs and low synchrotron peaked sources (LSP) \citep[e.g.,][and references therein]{boettcher13}. Recent studies of optical and $\gamma$-ray variability in a large sample of HSP and LSP blazars \citep{hovatta14} well agrees with above jet model. The latter also used to reproduce the radio-optical-$\gamma$-ray correlations to show that the variations of the jet power (or accretion power) is able to reproduce the observed multi-band trends \citep{arshakian12}. However, the physical models depend on distance of high-energy production region from a black hole. It is still not clear where and by mechanism the high energy photons are generated. In particularly, the source of external photon field for the EC mechanism can be the radiation from the disk, BLR,  molecular torus or from an hot corona \citep{sikora94,dermer93}. 
Milti-band monitorings of $\gamma$-ray blazars show that $\gamma$-ray emission can be generated in the disk, in the innermost part of the jet around radio core and/or RCS(s), and downstream of the RCS (see also discussions in \cite{tavares11}). We further discuss studies of the $\gamma$-ray production beyond the radio core. \cite{schinzel12} found from the VLBA, optical and $\gamma$-ray monitoring of the quasar 3C 345 that $\gamma$-ray emission from 0.1-300 GeV is driven by relativistic outflows of
the jet and is produced in a region of the jet that extends up to $\sim 23$ pc. The $\gamma$-ray emission is proposed to arise from SSC. Location of the $\gamma$-ray flare in the jet of BL Lac OJ 287 is reported by Agudo et al. (2011). They studied the $\gamma$-ray and 7 $mm$-wave
 flares and found that they are co-spatial and occur at 14 pc from the central
engine. The SSC scattering of optical photons from the flares is the likely mechanism of $\gamma$-ray generation.
Based on coincidence of the brightest $\gamma$-ray events and initial stages of a $mm$ flare in blazars,  \cite{tavares11} argued that the radio core or more likely downstream of it  (the RCS?) is the likely region of $\gamma$-ray production. It is located at about seven parsecs along the jet and well beyond the broad-line region (BLR). Likely generation mechanisms of  $\gamma$-ray emission are SSC from optical photons of the jet or EC from seed photons of the outflowing BLR.

\section{Jet-excited BLR: additional source of external photon field}
There is growing number of evidence for a jet-excited, outflowing gas in blazars. 
Detailed analyses of the line variability \citep[Ly$\alpha$ and C IV;][]{paltani03} of 3C 273 evidenced for two components characterized by virialized motion and by high, blueshifted velocities, which is interpreted as an outflowing, collisionally excited gas heated by the infrared radiation of the jet.
Strong manifestation of outflowing BLR clouds towards the line-of-sight of the observer is the blueshifted central component ($\sim -350$ km s$^{-1}$) of the H$\beta$ broad emission line in 3C 390.3 \citep{arshakian08}. The central blushifted peak appears during the high state of jet activity and disappears when the accretion power of the disk is dominant.
Existence of the jet-excited BLR outflowing downstream the jet was proposed to explain the link between jet kinematics on subpc-scales, optical continuum and emission line variability 
\citep{arshakian10,tavares10}. Non-thermal optical emission from the relativistic flow is beamed downstream the jet direction and may excite the surrounding gas on different spatial scales from the central black hole. The distance of the radio core from a black hole is not well known. If the radio core is located in the central engine, between the black hole and the virialized BLR, and it is optically thick to synchrotron emission then the high frequency emission zones (from optical to $\gamma$-ray) should appear upstream with respect to the radio core because of synchrotron opacity. This scenario is supported by finding for a sample of 183 blazars that the $\gamma$-ray emission (0.1-100 GeV) leads the radio emission (15 GHz) with time delay 1-8 months \citep{pushkarev10}. Downstream from the radio core, the optical and $\gamma$-ray production regions can be generated by SSC and EC (of optical photons from accretion disk or BLR) mechanisms. In either case, beamed optical emission can excite the virialized BLR as well as the outflowing BLR in the immediate vicinity of the jet. (Note that the outflowing BLR is related to a relativistic jet rather than to a virialized BLR and can manifest itself at distances far beyond the virialized BLR.) If the radio core is located beyond the virialized BLR then the beamed optical emission of the jet excites only the outflowing BLR. 

Direct observational evidence of the BLR close to the radio core of the jet comes from a 
response of the broad emission lines to changes of the non-thermal continuum emission of the jet \citep{tavares13}. The highest levels of the emission {Mg\,II} line flux and $\gamma$-ray outburst happens when a jet component passes through (or ejected from) the radio core of 3C\,454.3. This remarkable event was also confirmed in consequent studies \citep{isler13,jorstad13}.

Location of the radio core (and, hence, the outflowing BLR) can be established from the 
absorption signatures of $\gamma$-ray emission due to the interaction with radiation field of the virialized BLR \citep{pacciani13}. Almost all ten FSRQs of their sample have negligible or no evidence of absorption suggesting that the $\gamma$-ray production region is located beyond the virialized BLR. If the outflowing BLR exists around the radio core why it does not produce $\gamma$-ray absorption features? One reason could be that the jet excites a small region of the outflowing BLR around the radio core and the path length of the $\gamma\gamma$ scattering with the photons is short and the absorption is not significant. Another possibility is that the filling factor of the gas is low. And lastly, the visibility of absorption features can depend on the viewing angle and opening angle of the outflowing BLR. The $\gamma\gamma$ scattering will be stronger for close orientations of the outflowing BLR to the line of sight, while at angles larger than the half opening angle the scattering of $\gamma\gamma$ emission by outflowing BLR photons will be negligible. Combination of these three factors are also possible.  

The origin of the jet-excited outflow is most likely to be the wind from the accretion disk accelerated to sub-relativistic speeds by twisted magnetic field of the jet arising from the accretion disk. Relativistic flow may also interact with virialized BLR clouds or a star orbiting close to the central black hole and drag a gas downstream the jet. The latter was invoked to explain the short time-scale GeV flare-like event generated close to the radio core in 3C 454.3 \citep{khangulyan13}.

\section{Summary}
The innermost region of the jet, between the base of the jet, the radio core and RCS, in blazars
is the site of generation of the bulk of non-thermal variable emission across the entire electromagnetic spectrum. The jet-excited outflowing BLR (i) can contribute significantly to the emission line profile when the jet activity is at maximum and the jet central axis oriented closer to the line of sight and (ii) may serve as a source of high energy production along the parsec-scale jet by providing the optical seed photons 
for generating the high energy emission by means of EC. The fact that additional EC mechanism is required to fit the SED of FSRQs and LSP sources suggests that, in these blazars, the gas from the accretion disk effectively outflows and accelerates along the jet or dragged from virialized BLR by relativistic magnetized jet flow and excited by beamed optical emission of the jet. 

Absence of $\gamma$-ray absorption features in FSRQs suggests that the jet-excited region of the outflowing BLR is small and/or the density of optical photons is low in the excited region, and that the orientation of the outflowing BLR with respect to the line of sight 
is important for shaping the $\gamma$-ray absorption. 

%

%
%

\end{document}